\documentstyle[12pt,epsfig]{article}
\newcommand{\bea}{\begin{eqnarray}}
\newcommand{\eea}{\end{eqnarray}}
\newcommand{\be}{\begin{equation}}
\newcommand{\ee}{\end{equation}}
\newcommand{\nn}{\nonumber}
\newcommand{\rf}[1]{(\ref{#1})}

\begin{document}


\begin{center}
{\large\bf Mesons of the $f_0$ family in processes
$\pi\pi\to\pi\pi,K\overline{K}$ up to 2 GeV and chiral symmetry
}\footnote{This report has been supported by
the Grant Program of Plenipotentiary of Slovak Republic at JINR. Yu.S. and
M.N. were supported in part by the Slovak Scientific Grant Agency, Grant
VEGA No. 2/7175/20; and D.K., by Grant VEGA No. 2/5085/99.}\\
\bigskip
{\bf Yu.S.~Surovtsev}$^{a,}$\footnote{E-mail address: surovcev@thsun1.jinr.ru},
{\bf D.~Krupa}$^b$\footnote{E-mail address: krupa@savba.sk},
{\bf M.~Nagy}$^b$\footnote{E-mail address: fyzinami@nic.savba.sk}
\bigskip

$^a${\it Bogoliubov Laboratory of Theoretical Physics, Joint Institute for
Nuclear Research, Dubna 141 980, Moscow Region, Russia},\\
$^b${\it Institute of Physics, Slov.Acad.Sci., D\'ubravsk\'a cesta 9,
842 28 Bratislava, Slovakia}
\end{center}

\begin{abstract}
In a combined analysis of the experimental data on the coupled processes
$\pi\pi\to\pi\pi,K\overline{K}$ in the channel with $I^GJ^{PC}=0^+0^{++}$,
the various scenarios of these reactions (with different numbers of
resonances) are considered. In a model-independent approach, based only on
analyticity and unitarity, a resonance is represented by of a pole cluster
(poles on the Riemann surface) of the definite type that is defined by the
state nature. The best scenario contains the resonances $f_0(665)$ (with
properties of the $\sigma$-meson), ${f_0}(980)$ (with a dominant $s{\bar s}$
component), $f_0(1500)$ (with a dominant flavour-singlet {\it e.g.}, glueball
component) and the ${f_0}(1710)$ (with a considerable $s{\bar s}$ component).
If the ${f_0}(1370)$ exists, it has a dominant $s{\bar s}$ component. The
coupling constants of observed states with the considered channels and the
$\pi\pi$ and $K\overline{K}$ scattering lengths are obtained.  The conclusion
on the linear realization of chiral symmetry is drawn.
\end{abstract}

\section{Introduction}
In the scalar mesonic sector, many states have been discovered at present
\cite{PDG-00}, however, their assignment to quark-model configurations is
problematic -- one can compare various variants of that assignment,
for example, \cite{Torn}-\cite{Volkov-Yudichev}.
It seems that the problem of scalar mesons is far off the solution up to now.
For instance, at present, additional arguments have been added by
N.N.~Achasov \cite{Achasov00} in favour of the 4-quark nature of
${f_0}(980)$ and
${a_0}(980)$ mesons on the basis of interpretation of the experimental data
on the decays $\phi\to\gamma\pi^0\pi^0,\gamma\pi^0\eta$ \cite{Novosib}.
On the other hand, F.E.~Close and A.~Kirk \cite{Close-Kirk} have shown that
mixing between the $f_0(980)$ and $a_0(980)$ radically affects some existing
predictions of their produciton in $\phi$ radiative decay. Generally, the
4-quark interpretation, beautifully solving the old problem of the unusual
properties of scalar mesons, sets new questions.
Where are the 2-quark states, their radial excitations and the other members
of 4-quark multiplets $9,9^*,36$ and $36^*$, which are predicted to exist
below 2.5 GeV \cite{Jaffe}?

The discovered states in the scalar sector and their properties do not allow
one to make up the scalar $q{\bar q}$ nonet and to solve other exiting
questions up to now. Generally, difficulites in understanding the
scalar-isoscalar sector seem to be related to both the hard-accounting
influence of the vacuum (and such effects as the instanton contributions) and
a strong model dependence of an information about wide multichannel states.

Earlier, we have shown \cite{KMS-nc96} that an inadequate description of
multichannel states (to which scalar mesons belong) gives not only their
distorted parameters when analyzing data but also can cause the fictitious
states when one neglects important (even energetic-closed) channels.
Obviously, it is important to have a model-independent information on
investigated states and on their QCD nature. It can be obtained only on the
basis of the first principles (analyticity, unitarity) immediately applied to
experimental data analysis. Earlier, we have proposed this method for 2- and
3-channel resonances and developed the concept of standard clusters (poles on
the Riemann surface) as a qualitative characteristic of a state and a
sufficient condition of its existence \cite{KMS-nc96}. We outline this below
for the 2-channel case of the coupled processes
$\pi\pi\to\pi\pi,K\overline{K}$ in the channel with $I^GJ^{PC}=0^+0^{++}$.
Since, in this work, a main stress is laid on studying lowest states, it is
sufficient to restrict oneself to a two-channel approach when considering
simultaneously the coupled processes $\pi\pi\to\pi\pi,K\overline{K}$, though
in the future it is nesessary to take into account the thresholds of other
coupled processes, first of all, of $\eta\eta$ and $\eta\eta^{\prime}$
scatterings.
In this work, we are going to show that the large background, which one has
obtained earlier in various analyses of the $s$-wave $\pi\pi$ scattering
\cite{PDG-00}, hides, in reality, the $\sigma$-meson \cite{NJL} below 1 GeV
and the effect of the left-hand branch-point. Therefore, in the uniformizing
variable, one must take into account, besides the branch-points corresponding
to the thresholds of the processes $\pi\pi\to\pi\pi,K\overline{K}$, also the
left-hand branch-point at $s=0$, related to the background in which the
crossing-channel contributions are contained \cite{PRD-01}.
Furthermore, we shall obtain definite indications of the QCD nature of other
$f_0$ resonances and of the linear realization of chiral symmetry.

Note that recent new analyses of old and new experimental data found a
candidate for the $\sigma$-meson below 1 GeV (see, {\it e.g.}, \cite{Torn},
\cite{Anisovich}, \cite{Zou}-\cite{Li-Zou-Li}).
However, these analyses use either the Breit -- Wigner form that is
insufficiently-flexible even if modified or the $K$-matrix formalism without
taking account of the energetic-closed (maybe, important) channels, or
specific forms of interactions in the quark models; therefore, there one
cannot talk about the model independence of results. Besides, in these analyses,
a large $\pi\pi$-background is obtained.

The layout of this paper is as follows. In Section II, we outline the
two-coupled-channel formalism, determine the pole clusters on the Riemann
surface as characteristics of multichannel states, and introduce a new
uniformizing variable, allowing for the branch-points of the right-hand
(unitary) and left-hand cuts of the $\pi\pi$-scattering amplitude.
In Section III, we analyze simultaneously experimental data on the processes
$\pi\pi\to \pi\pi,K\overline{K}$ in the isoscalar $s$-wave on the basis of
the presented approach. Note that earlier we have shown in two approaches
\cite{KMS-nc96} and \cite{PRD-01} without and with taking into account the
left-hand branch-point $\sqrt{s}$ in the uniformizing variable, respectively,
that the minimal scenario of the simultaneous description of two coupled
processes $\pi\pi\to\pi\pi,K\overline{K}$ is realized without the $f_0(1370)$
and $f_0(1710)$ that are in the Particle Data Group tables. Therefore,
we consider also the variants of description of the indicated processes
including these states separately as well as simultaneously, and obtain
indications of their QCD nature different from the results of many other
works (note that our approach is based on the first principles and is free
from dynamic assumtions, therefore, our results are rather
model-independent). In the Conclusion, the obtained results are discussed.

\section{Two-Coupled-Channel Formalism}

Here we restrict ourselves to a 2-channel consideration of the coupled
processes $\pi\pi\to \pi\pi,K\overline{K}$. Therefore, we have the 2-channel
$S$-matrix determined on the 4-sheeted Riemann surface. The matrix elements
$S_{\alpha\beta}$, where $\alpha,\beta=1(\pi\pi), 2(K\overline{K})$, have the
right-hand cuts along the real axis of the $s$-plane, starting at $4m_\pi^2$
and $4m_K^2$, and the left-hand cuts, beginning at $s=0$ for $S_{11}$ and at
$4(m_K^2- m_\pi^2)$ for $S_{22}$ and $S_{12}$. The Riemann-surface sheets are
numbered according to the signs of analytic continuations of the channel
momenta
\be
k_1=(s/4-m_\pi^2)^{1/2}, ~~~~~~~~k_2=(s/4-m_K^2)^{1/2}
\ee
as follows:  ~ signs $({\mbox{Im}}k_1,{\mbox{Im}}k_2)=++,-+,--,+-$ correspond
to the sheets I, II, III, IV.

To obtain the resonance representation on the Riemann surface, we express
analytic continuations of the matrix elements to the unphysical sheets
$S_{\alpha\beta}^L$ ($L=II,III,IV$) in terms of those on the physical sheet
$S_{\alpha\beta}^I$:
\bea \label{S_L}
&&S_{11}^{II}=\frac{1}{S_{11}^I},\qquad
~~~~S_{11}^{III}=\frac{S_{22}^I}{\det S^I},
\qquad S_{11}^{IV}=\frac{\det
S^I}{S_{22}^I},\nn\\ &&S_{22}^{II}=\frac{\det
S^I}{S_{11}^I},\qquad
S_{22}^{III}=\frac{S_{11}^I} {\det S^I},\qquad
S_{22}^{IV}=\frac{1}{S_{22}^I},\\
&&S_{12}^{II}=\frac{iS_{12}^I}{S_{11}^I},\qquad
~~~S_{12}^{III}=\frac{-S_{12}^I} {\det
S^I},\qquad
S_{12}^{IV}=\frac{iS_{12}^I}{S_{22}^I},\nn
\eea
Here $\det S^I=S_{11}^I S_{22}^I-(S_{12}^I)^2$;
$(S_{12}^I)^2=-s^{-1}\sqrt{(s-4m_\pi^2)(s-4m_K^2)}F(s)$; in the limited
energy interval, $F(s)$ is proportional to the squared product of the
coupling constants of the considered state with channels 1 and 2. These
formulas are convenient by that the $S$-matrix elements on the physical sheet
$S_{\alpha\beta}^I$ have, except for the real axis, only zeros corresponding
to resonances, at least, around the physical region that is interesting
for us.
Formulas \rf{S_L} immediately give the resonance representation by poles and
zeros on the 4-sheeted Riemann surface. One must distinguish between three
types of 2-channel resonances described by a pair of conjugate zeros on
sheet I:
({\bf a}) in $S_{11}$, ({\bf b}) in $S_{22}$, ({\bf c}) in each of $S_{11}$
and $S_{22}$. As seen from eqs. \rf{S_L}, to the resonances of types
({\bf a}) and ({\bf b}), there corresponds a pair of complex conjugate poles
on sheet III shifted relative to a pair of poles on sheet II and IV,
respectively. For the states of type ({\bf c}), one must consider the
corresponding two pairs of conjugate poles on sheet III. A resonance of every
type is represented by a pair of complex-conjugate clusters (of poles and
zeros on the Riemann surface) of a size typical of strong interactions. The
cluster kind is related to the state nature. The resonance coupled relatively
more strongly to the $\pi\pi$ channel than to the $K\overline{K}$ one is
described by the cluster of type ({\bf a}); in the opposite case, it is
represented by the cluster of type ({\bf b}) (say, the state with the
dominant $s{\bar s}$ component); the flavour singlet ({\it e.g.} glueball)
must be represented by the cluster of type ({\bf c}) as a necessary
condition, if this state lies above the thresholds of considered channels.

Furthermore, according to the type of pole clusters, we can distinguish, in a
model-independent way, a bound state of colourless particles ({\it e.g.},
$K\overline{K}$ molecule) and a $q{\bar q}$ bound state \cite{KMS-nc96,MP-93}.
Just as in the 1-channel case, the existence of a particle bound-state means
the presence of a pole on the real axis under the threshold on the physical
sheet, so in the 2-channel case, the existence of a particle bound-state in
channel 2 ($K\overline{K}$ molecule) that, however, can decay into channel
1 ($\pi\pi$ decay), would imply the presence of a pair of complex conjugate
poles on sheet II under the second-channel threshold without an accompaniment
of the corresponding shifted pair of poles on sheet III. Namely, according
to this test, earlier, the interpretation of the $f_0(980)$ state as a
$K\overline{K}$ molecule has been rejected.

For the simultaneous analysis of experimental data on coupled processes, it
is convenient to use the Le Couteur-Newton relations \cite{LC} expressing
the $S$-matrix elements of all coupled processes in terms of the Jost matrix
determinant $d(k_1,k_2)$, the real analytic function with the only
square-root branch-points at $k_i=0$. To take into account, in addition to
the latter, also the left-hand branch-point at $s=0$, the uniformizing
variable is used \footnote{The analogous uniformizing variable has
been used, {\it e.g.}, in Ref. \cite{Meshch} at studying the forward elastic
$p{\bar p}$ scattering amplitude.}
\be \label{v}
v=\frac{m_K\sqrt{s-4m_\pi^2}+m_\pi\sqrt{s-4m_K^2}}{\sqrt{s(m_K^2-m_\pi^2)}}.
\ee
It maps the 4-sheeted Riemann surface with two unitary cuts and the
left-hand cut onto the $v$-plane. (Note that other authors have used the
parameterizations with the
Jost functions at analyzing the $s$-wave $\pi\pi$ scattering in the
one-channel approach \cite{Bohacik} and in the two-channel one \cite{MP-93}.
In latter work, the uniformizing variable $k_2$ has been used, therefore,
their approach cannot be employed near by the $\pi\pi$ threshold.)
\begin{figure}[ht]
\centering
\epsfxsize=9cm\epsffile{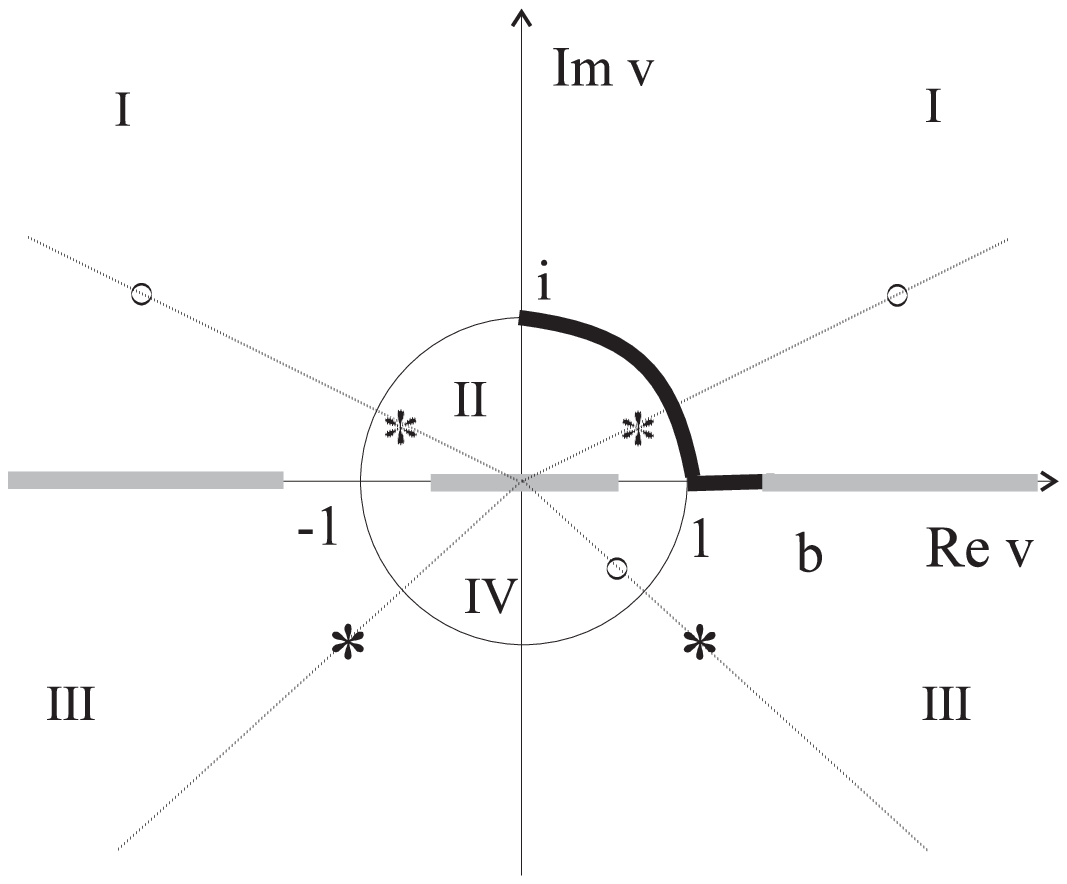}
\vskip -.2cm
\caption{Uniformization plane for the \protect$\pi\pi$-scattering amplitude.}
\label{fig:v.plane}
\end{figure}
In Fig.\ref{fig:v.plane}, the plane of the uniformizing variable $v$ for the
$\pi\pi$-scattering amplitude is depicted. The Roman numerals (I,\ldots, IV)
denote the images of the corresponding sheets; the thick line represents
the physical region; the points i, 1 and $b=\sqrt{(m_K+m_\pi)/(m_K-m_\pi)}$
correspond to the $\pi\pi, K\overline{K}$ thresholds and $s=\infty$,
respectively; the shaded intervals
$(-\infty,-b],~[-b^{-1},~b^{-1}],~[b,\infty)$ are the images of the
corresponding edges of the left-hand cut. The depicted positions of poles
($*$) and of zeros ($\circ$) give the representation of the type ({\bf a})
resonance in $S_{11}$.

On the $v$-plane, $S_{11}$ has no cuts; however, $S_{12}$ and $S_{22}$ do
have the cuts which arise from the left-hand cut on the $s$-plane, starting
at $s=4(m_K^2-m_\pi^2)$, which further is neglected in the Riemann-surface
structure, and the contribution of this cut is taken into account in the
$K\overline{K}$ background as a pole on the real $s$-axis on the physical
sheet in the sub-$K\overline{K}$-threshold region.

On $v$-plane, the Le Couteur-Newton relations are \cite{KMS-nc96,LC}
\be \label{v:C-Newton}
S_{11}=\frac{d(-v^{-1})}{d(v)},\quad
S_{22}=\frac{d(v^{-1})}{d(v)},
\quad S_{11}S_{22}-S_{12}^2=\frac{d(-v)}{d(v)}.
\ee
The $d(v)$-function already does not possess branch-points and is taken as
\be
d=d_B d_{res},
\ee
where ~$d_B=B_{\pi}B_K$;
$B_{\pi}$ contains the possible remaining $\pi\pi$-background contribution,
related to exchanges in crossing channels (the consequent analysis gives
$B_{\pi}=1$); $B_K$ is that part of the $K\overline{K}$ background which does
not contribute to the $\pi\pi$-scattering amplitude:
\be \label{B_K}
B_K=v^{-4}(1-v_0v)^4(1+v_0^*v)^4.
\ee
The fourth power in (\ref{B_K}) is stipulated by the following
model-independent arguments \cite{PRD-01}. First, a pole on the real $s$-axis
on the physical sheet in $S_{22}$ is accompanied by a pole in sheet II at the
same $s$-value (as seen from eqs.(\ref{S_L})). On the $v$-plane this implies
the pole of second order (and also zero of the same order, symmetric to the
pole with respect to the real axis). Second, for the $s$-channel process
$\pi\pi\to K\overline{K}$, the crossing $u$- and $t$-channels are the $\pi-K$
and $\overline{\pi}-K$ scattering (exchanges in these channels give
contributions on the left-hand cut). This results in the additional doubling
of the multiplicity of the indicated pole on the $v$-plane. So, the model of
the $K\overline{K}$ background is determined by poles (here by the single
one) on the real $s$-axis at the left-hand cut position on the physical sheet.

The function $d_{res}(v)$ represents the contribution of resonances,
described by one of three types of the pole-zero clusters, {\it i.e.},
\be \label{d_res}
d_{res} = v^{-M}\prod_{n=1}^{M} (1-v_n^* v)(1+v_n v),
\ee
where $M$ is the number of pairs of the conjugate zeros.

\section{Analysis of experimental data}

We simultaneously analyze the available experimental data on the
$\pi\pi$-scattering \cite{Hyams} and the process $\pi\pi\to K\overline{K}$
\cite{Wetzel} in the channel with $I^GJ^{PC}=0^+0^{++}$. As the data, we use
the results of phase analyses which are given for phase shifts of the
amplitudes ($\delta_1$ and $\delta_{12}$) and for moduli of the $S$-matrix
elements $\eta_1=|S_{aa}|$ (a=1$\pi\pi$,2$K\overline{K}$) and $\xi=|S_{12}|$.
The 2-channel unitarity condition gives ~~$\eta_1=\eta_2=\eta, ~~
\xi=(1-\eta^2)^{1/2},~~ \delta_{12}=\delta_1+\delta_2$.

We consider four variants, in which the following states are taken into
account: \\
\underline{Variant 1}: The $f_0(665)$ and $f_0 (980)$) with the clusters of
type ({\bf a}), and $f_0(1500)$, of type ({\bf c}); \\
\underline{Variant 2}: The same three resonances $+$ the ${f_0}(1370)$ of
type ({\bf b});\\
\underline{Variant 3}: The $f_0(665)$, $f_0 (980)$) and $f_0 (1500)$ $+$ the
${f_0}(1710)$ of type ({\bf b});\\
\underline{Variant 4}: All the five resonances of the indicated types.

\begin{table}[ht]
\centering \caption{}
\vskip0.3truecm
\begin{tabular}{|c|cc|cc|cc|cc|cc|cc|cc|cc|}
\hline {Variant} &
\multicolumn{4}{c|}{1}& \multicolumn{4}{c|}{2}\\
\hline Quantity &
$\delta_1$ & $\eta$ & $\delta_{12}$ & $\xi$ &
$\delta_1$ & $\eta$ &
$\delta_{12}$ & $\xi$ \\ \hline Number of & 160
& 50 & 80 & 33 & 160 & 50
& 80 & 39\\ exp.points & {}& & {}& & {}& & {}&
\\ \hline
$\chi^2/\mbox{N}_{DF}$ & 2.7 & 0.72 & 2.37 & 1.1
& 2.85 & 0.82 & 3.98 &
0.92\\ \hline $\chi^2/\mbox{N}_{DF}$ &
\multicolumn{2}{c|}{1.96} &
\multicolumn{2}{c|}{2.01} &
\multicolumn{2}{c|}{2.01} &
\multicolumn{2}{c|}{3} \\ \hline
$\chi^2/\mbox{N}_{DF}$ &
\multicolumn{4}{c|}{1.98} &
\multicolumn{4}{c|}{2.45}\\ \hline $v_0$ &
\multicolumn{4}{c|}{$0.954381+0.29859i$} &
\multicolumn{4}{c|}{$0.97925+0.202657i$} \\
$(s_0,{\rm GeV}^2)$ &
\multicolumn{4}{c|}{(0.441)} &
\multicolumn{4}{c|}{(0.6466)} \\ \hline
\hline {Variant} & \multicolumn{4}{c|}{3} &
\multicolumn{4}{c|}{4} \\
\hline Quantity & $\delta_1$ & $\eta$ &
$\delta_{12}$ & $\xi$ & $\delta_1$
& $\eta$ & $\delta_{12}$ & $\xi$ \\ \hline
Number of & 160 & 50 & 80 & 42
& 160 & 50 & 80 & 42 \\ exp.points & {} & & {} &
& {} & & {} & \\ \hline
$\chi^2/\mbox{N}_{DF}$ & 2.38 & 0.8 & 2.25 &
0.92 & 2.57 & 0.85 & 4.74 &
1.05\\ \hline $\chi^2/\mbox{N}_{DF}$&
\multicolumn{2}{c|}{1.72} &
\multicolumn{2}{c|}{1.8} &
\multicolumn{2}{c|}{1.81} &
\multicolumn{2}{c|}{3.49}\\ \hline
$\chi^2/\mbox{N}_{DF}$&
\multicolumn{4}{c|}{1.76} &
\multicolumn{4}{c|}{2.59}\\ \hline $v_0$ &
\multicolumn{4}{c|}{$0.954572+0.29798i$} &
\multicolumn{4}{c|}{$0.982091+0.188405i$}\\
$(s_0,{\rm GeV}^2)$ &
\multicolumn{4}{c|}{(0.4646)} &
\multicolumn{4}{c|}{(0.678)}\\ \hline
\end{tabular} \label{tab:quality} \end{table}

%
\begin{figure}[ht]
\centering
\begin{tabular}{cc}
\epsfxsize=8cm\epsffile{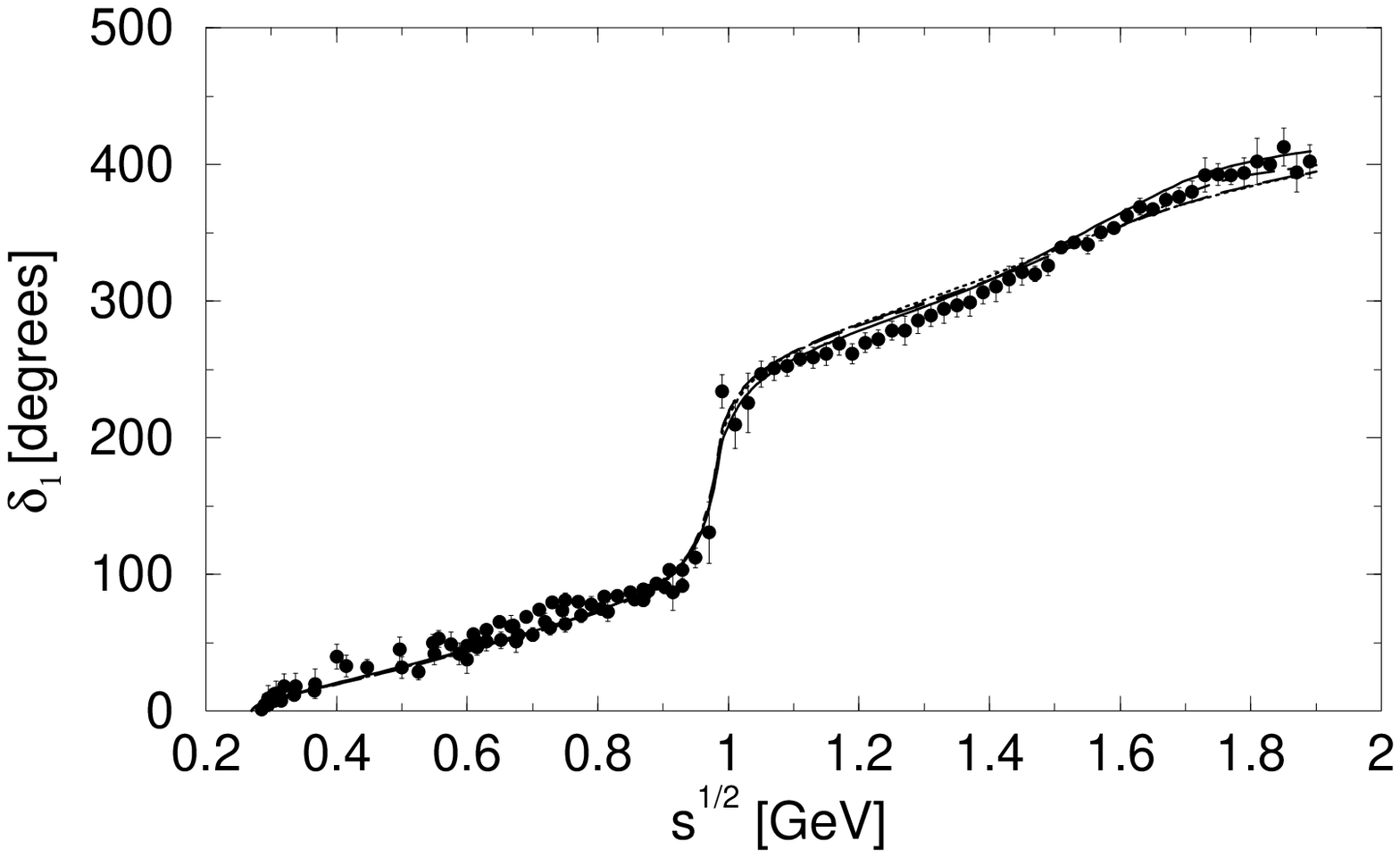}&
\epsfxsize=7.8cm\epsffile{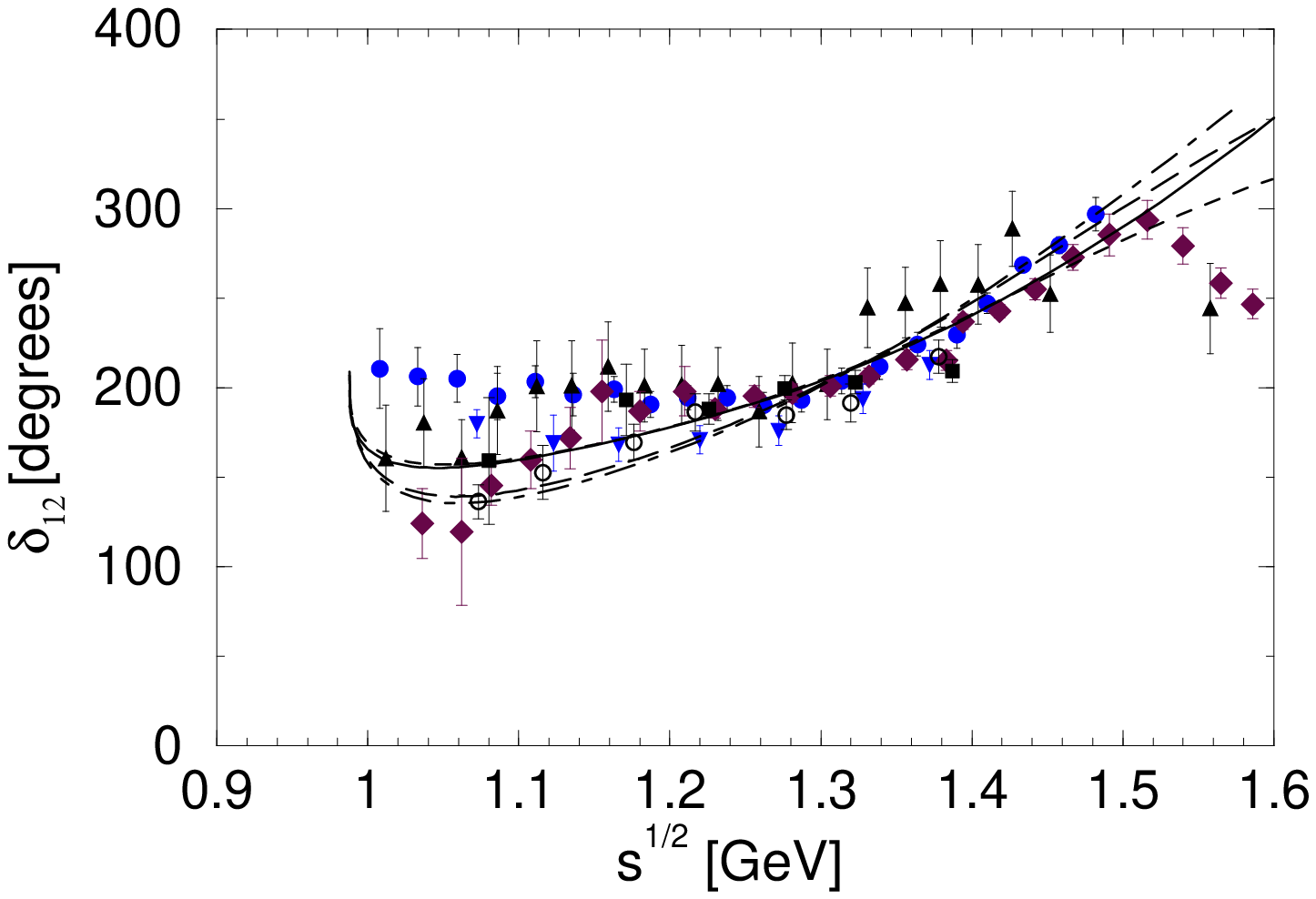}\\
\epsfxsize=7.8cm\epsffile{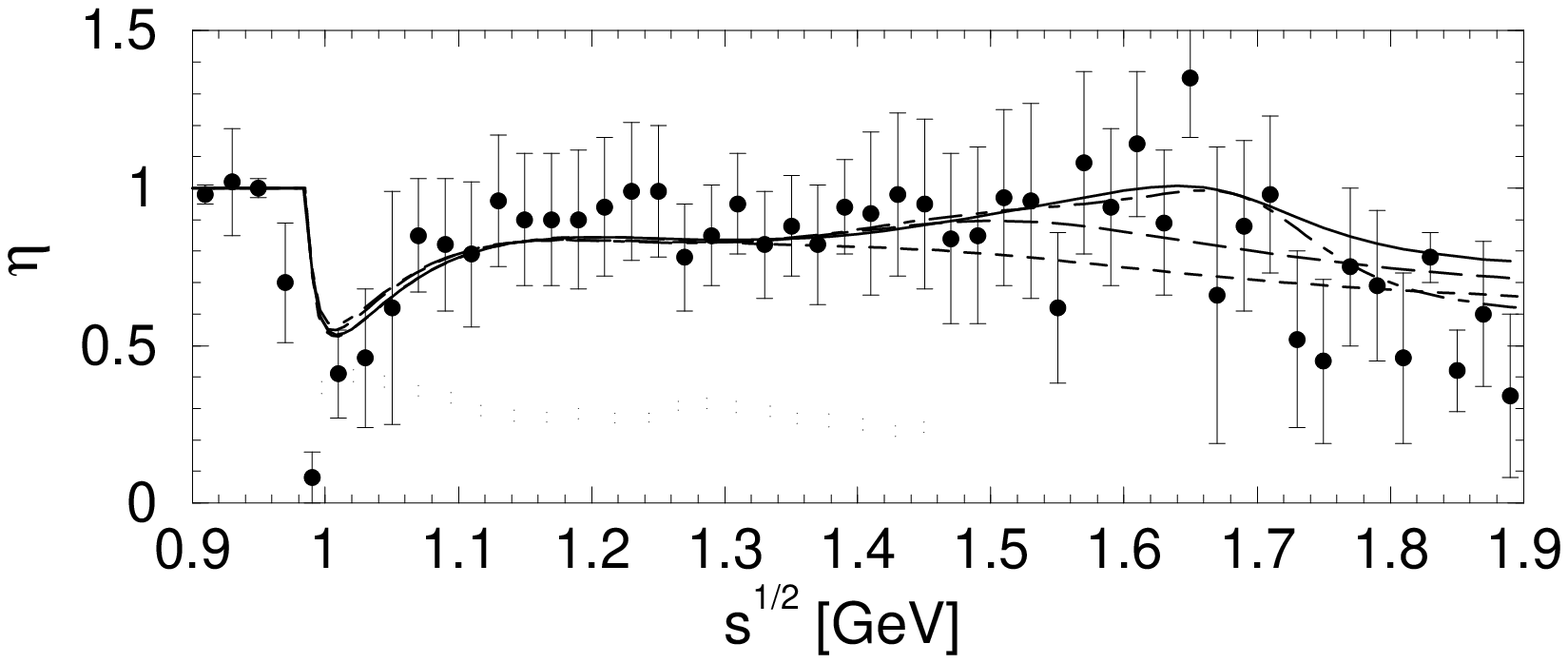}&
\epsfxsize=8cm\epsffile{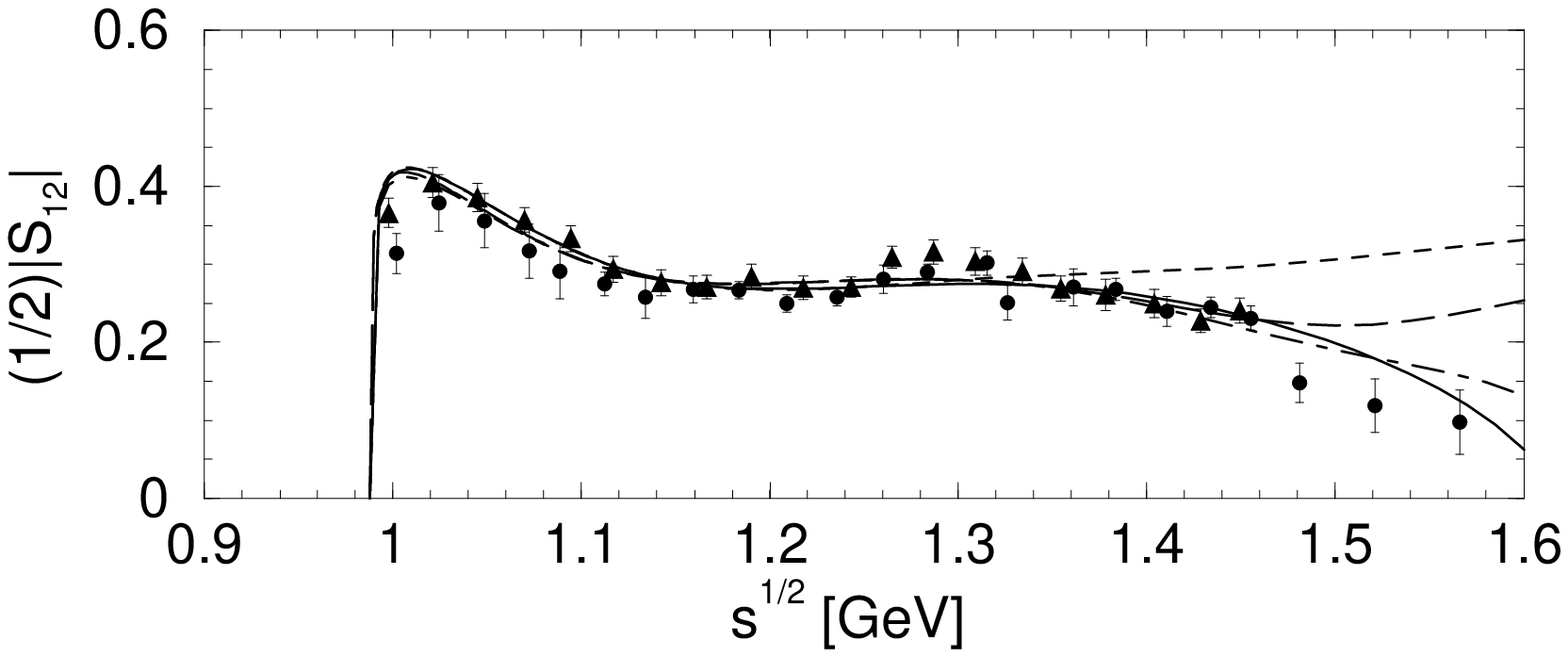}\\
\end{tabular}
\vskip -.3cm
\caption{ }
\label{fig:quantities}
\end{figure}

The other possibilities of the representation of these states are rejected
by our analysis. We consider these variants, because the minimal possibility
of the simultaneous description of two coupled processes $\pi\pi\to\pi\pi,
K\overline{K}$ is realized without the $f_0(1370)$ and $f_0(1710)$, as it is
shown in our work \cite{PRD-01}.
The $\pi\pi$-scattering data are described from the threshold
to 1.89 GeV in all the four variants (in addition, we take $B_\pi=1$) and
are taken from the analysis by B. Hyams {\em et al.} \cite{Hyams} in this
energy region, and below 1 GeV, from many works \cite{Hyams}. For the
reaction $\pi\pi\to K\overline{K}$, practically all the accessible data are
used, but the description ranges are slightly different for various variants
and extend from the threshold to $\sim$ 1.4 GeV for variant 1, to $\sim$
1.46 GeV for variant 2, and to $\sim$ 1.5 GeV for variants 3 and 4.
Table~\ref{tab:quality} demonstrates the
quality of fits to the experimental data (the number of fitted parameters is
17 for variant 1, 21 for variants 2 and 3, 25 for variant 4).
When calculating $\chi^2/\mbox{N}_{DF}$, we have rejected the experimental
points at 0.61, 0.65, and 0.73 GeV for $\delta_1$, at 0.99, 1.65, and 1.85
GeV for $\eta$, at 1.111, 1.163, and 1.387 GeV for $\delta_{12}$, and at
1.002, 1.265, and 1.287 GeV for $\xi$ that give an especially large
contribution to $\chi^2$. We note that two variants (1 and 3) are the best,
both without the ${f_0}(1370)$, and we stress that this analysis uses the
{\it parameterless} description of the $\pi\pi$ background.

Let us indicate the obtained zero positions, on the $v$ plane, of the
corresponding resonances:\\
\underline{Variant 1}:
\bea
{\rm for} f_0(665):~~~~~~&v_1=1.36964+0.208632i,~~~~&v_2=0.921962-0.25348i,\nn\\
{\rm for} f_0(980):~~~~~~&v_3=1.04834+0.0478652i,~~~&v_4=0.858452-0.0925771i,\nn\\
{\rm for} f_0(1500):~~~~~&v_5=1.2587+0.0398893i,~~~~&v_6=1.2323-0.0323298i,\nn\\
                    &v_7=0.809818-0.019354i,~~&v_8=0.793914-0.0266319i,\nn
\eea
\underline{Variant 2}:
\bea
{\rm for} f_0(665):~~~~~~&v_1=1.36783+0.212659i,~~~~&v_2=0.921962-0.25348i,\nn\\
{\rm for} f_0(980):~~~~~~&v_3=1.04462+0.0479703i,~~~&v_4=0.858452-0.0925771i,\nn\\
{\rm for} f_0(1370):~~~~~&v_5=1.22783-0.0483842i,~~~&v_6=0.802595-0.0379537i,\nn\\
{\rm for} f_0(1500):~~~~~&v_7=1.2587+0.0398893i,~~~~&v_8=1.24837-0.0358916i,\nn\\
                         &v_9=0.804333-0.0179899i,~~&v_{10}=0.795579-0.0253985i,\nn
\eea
\underline{Variant 3}:
\bea
{\rm for} f_0(665):~~~~~~&v_1=1.38633+0.230588i,~~~~&v_2=0.904085-0.263033i,\nn\\
{\rm for} f_0(980):~~~~~~&v_3=1.05103+0.0487473i,~~~&v_4=0.864109-0.0922272i,\nn\\
{\rm for} f_0(1500):~~~~~&v_5=1.2477+0.0321349i,~~~~&v_6=1.24027-0.0384191i,\nn\\
                    &v_7=0.804333-0.0179899i,~~&v_8=0.795579-0.0253985i,\nn\\
{\rm for} f_0(1710):~~~~~&v_9=1.25928-0.0115127i,~~~&v_{10}=0.795429-0.00629969i.\nn
\eea
\underline{Variant 4}:
\bea
{\rm for} f_0(665):~~~~~~&v_1=1.37103+0.21659i,~~~~~&v_2=0.917731-0.256026i,\nn\\
{\rm for} f_0(980):~~~~~~&v_3=1.0457+0.0507053i,~~~~&v_4=0.858452-0.0925771i,\nn\\
{\rm for} f_0(1370):~~~~~&v_5=1.22781-0.0496592i,~~~&v_6=0.801009-0.0420258i,\nn\\
{\rm for} f_0(1500):~~~~~&v_7=1.25817+0.0397054i,~~~&v_8=1.25078-0.0350769i,\nn\\
                         &v_9=0.809333-0.0179899i,~~&v_{10}=0.795579-0.0253985i,\nn\\
{\rm for} f_0(1710):~~~~~&v_{11}=1.2604-0.00927258i,&v_{12}=0.794694-0.00458088i.\nn
\eea

\begin{table}[ht]
\centering
\caption{ Pole clusters for considered resonances in variant 3. }
\vskip0.3truecm
\begin{tabular}{|c|c|c|c|c|}
\hline \multicolumn{2}{|c|}{Sheet} & II & III &IV \\
\hline
{$f_0 (665)$} & {E, MeV} & 570$\pm$13 & 700$\pm$15 & {}\\ {} &
{$\Gamma$, MeV} & 590$\pm$24 & 72$\pm$5 & {} \\
\hline {$f_0(980)$} & {E, MeV} & 989$\pm$5~ & 982$\pm$14 & {}\\ {} &
{$\Gamma$, MeV} & 29$\pm$7 & 195$\pm$21 & {} \\
\hline
{$f_0 (1500)$} &
{E, MeV} & 1505$\pm$23~ & 1490$\pm$30~~1510$\pm$25 & 1430$\pm$20~ \\ {} &
{$\Gamma$, MeV} & 272$\pm$25 & ~220$\pm$26~~~~370$\pm$30 & 275$\pm$32 \\
\hline
{$f_0 (1710)$} & {E, MeV} & {} & 1680$\pm$20~~ & 1700$\pm$15~ \\ {}
& {$\Gamma$, MeV} & {} & 114$\pm$15~ & 138$\pm$21
\\ \hline
\end{tabular} \label{tab:clusters3} \end{table}

\begin{table}[ht]
\centering
\caption{ Pole clusters for considered resonances in variant 4. }
\vskip0.3truecm
\begin{tabular}{|c|c|c|c|c|}
\hline \multicolumn{2}{|c|}{Sheet} & II & III &
IV \\ \hline {$f_0 (665)$}
& {E, MeV} & 600$\pm$16 & 715$\pm$17 & {}\\ {} &
{$\Gamma$, MeV} &
605$\pm$28 & 59$\pm$6 & {} \\ \hline {$f_0
(980)$} & {E, MeV} & 985$\pm$5~
& 984$\pm$18 & {}\\ {} & {$\Gamma$, MeV} &
27$\pm$8 & 210$\pm$22 & {} \\
\hline {$f_0 (1370)$} & {E, MeV} & {} &
1310$\pm$22~ & 1320$\pm$20~ \\ {}
& {$\Gamma$, MeV} & {} & 410$\pm$29 & 275$\pm$25
\\ \hline {$f_0 (1500)$}
& {E, MeV} & 1528$\pm$22~ &
1490$\pm$30~~1510$\pm$20 & 1510$\pm$21~ \\ {}
& {$\Gamma$, MeV} & 385$\pm$25 &
~220$\pm$24~~~~370$\pm$30 & 308$\pm$30 \\
\hline {$f_0 (1710)$} & {E, MeV} & {} &
1700$\pm$25~~ & 1700$\pm$20~ \\ {}
& {$\Gamma$, MeV} & {} & 86$\pm$16 & 115$\pm$20
\\ \hline \hline
\end{tabular} \label{tab:clusters4} \end{table}


Figures \ref{fig:quantities} demonstrate the comparison of the obtained
energy dependence of four analyzed quantities with the experimental data: The
short-dashed lines correspond to variant 1; the long-dashed curves, to
variant 2; the dot-dashed ones, to variant 4; and the solid lines, to
variant 3.

\begin{table}[htb] \centering \caption{ }
\vskip0.3truecm
\begin{tabular}{|c|c|c|c|c|} \hline {}  &
$f_0(665)$ & $f_0(980)$ &
$f_0(1370)$ & $f_0(1500)$\\ \hline $g_{1}$, GeV
& ~$0.652\pm 0.065$~ &
~$0.167 \pm 0.05$~~ & ~$0.116 \pm 0.03$~ &
~$0.657 \pm 0.113$~\\ \hline
$g_2$, GeV & ~$0.724\pm0.1$~~~~ &
~$0.445\pm0.031$~ & ~~$0.99\pm0.05$~ &
~$0.666\pm0.15$~~\\ \hline \end{tabular}
\label{tab:constants} \end{table}

In Tables~\ref{tab:clusters3} and \ref{tab:clusters4}, the obtained pole
clusters of considered resonances are shown on the corresponding sheets on
the complex energy plane ($\sqrt{s_r}={\rm E}_r-i\Gamma_r$) for the best
variant 3 (without the $f_(1370)$) and for variant 4 (with all five states).

The coupling constants of obtained states with $\pi\pi$ ($g_1$) and
$K\overline{K}$ ($g_2$) systems are calculated through the residues of
amplitudes at the pole on sheet II -- for resonances of types ({\bf a})
and ({\bf c}), and on sheet IV -- for resonances of type ({\bf b}).
Expressing the $T$-matrix via the $S$-matrix as
\be
S_{ii}=1+2i\rho_iT_{ii},~~~~~~S_{12}=2i\sqrt{\rho_1\rho_2} T_{12},
\ee
where $\rho_i=\sqrt{(s-4m_i^2)/s}$, and taking the resonance part of the
amplitude as
\be
T_{ij}^{res}=\sum_r g_{ir}g_{rj}D_r^{-1}(s)
\ee
with $D_r(s)$ being an inverse propagator ($D_r(s)\propto s-s_r$), we
show the results of that calculation in Table~\ref{tab:constants}.
We see that the $f_0(980)$ and especially the ${f_0}(1370)$ are coupled
essentially more strongly to the $K\overline{K}$ system than to the $\pi\pi$
one, which tells about the dominant $s{\bar s}$ component in these states.
The $f_0(1500)$ has the approximately equal coupling constants with the
$\pi\pi$ and $K\overline{K}$ systems, which apparently could point up to
its dominant glueball component \cite{Amsler-Close}. The coupling constant of
the $f_0(1710)$ with the $\pi\pi$ channel cannot be calculated by this
method, unless the description of $\pi\pi\to K\overline{K}$ reaction is
obtained in the region of this resonance. But this state is represented by
the cluster corresponding to the dominant $s{\bar s}$ component.

Let us also present the calculated scattering lengths: For the
$K\overline{K}$ scattering: \\
\hspace*{3.cm}$a_0^0 = -1.25\pm 0.11+(0.65\pm 0.09)i,~ [m_{\pi^+}^{-1}];
~~~~~~~~~~~~~({\rm variant 1}),$ \\
\hspace*{3.cm}$a_0^0= -1.548\pm 0.13+(0.634\pm 0.1)i,~ [m_{\pi^+}^{-1}];
~~~~~~~~~~~~({\rm variant 2}),$ \\
\hspace*{3.cm}$a_0^0 = -1.19\pm 0.08+(0.622\pm 0.07)i,~ [m_{\pi^+}^{-1}];
~~~~~~~~~~~~({\rm variant 3}),$\\
\hspace*{3.cm}$a_0^0 = -1.58\pm 0.12+(0.59\pm 0.1)i,~ [m_{\pi^+}^{-1}];
~~~~~~~~~~~~~~~({\rm variant 4}).$ \\
The presence of the imaginary part in $a_0^0(K\overline{K})$ reflects the
fact that already at the threshold of the $K\overline{K}$, other channels
($2\pi,4\pi$, etc.) are opened. Variants 2 and 4 include the $f_0(1370)$
unlike variants 1 and 3. We see that ${\rm Re}~a_0^0(K\overline{K})$ is very
sensitive to whether this state exists or not.

In Table~\ref{tab:pipi.length}, we compare our results for the $\pi\pi$
scattering length $a_0^0$ with results of some other theoretical and
experimental works.

\begin{table}[htb] \centering \caption{ }
\vskip0.3truecm
\begin{tabular}{|c|l|l|} \hline $a_0^0,
~m_{\pi^+}^{-1}$ &
~~~~~~~References & ~~~~~~~~~~~~~~~~~Remarks \\
\hline $0.27\pm 0.06$ (1)&
our paper & model-independent approach \\
$0.267\pm 0.07$ (2)&{}&{}\\
$0.28\pm 0.05$ (3)&{}&{}\\ $0.27\pm 0.08$
(4)&{}&{}\\ \hline $0.26\pm
0.05$ & L. Rosselet et al.\cite{Hyams} &
analysis of the decay $K\to\pi\pi
e\nu$ \\ {} & {} & using Roy's model\\ \hline
$0.24\pm 0.09$ & A.A.
Bel'kov et al.\cite{Hyams} & analysis of $\pi^-
p\to\pi^+\pi^-n$ \\ {} & {}
& using the effective range formula\\ \hline
$0.23$ & S. Ishida et
al.\cite{Ishida} & modified analysis of $\pi\pi$
scattering \\ {} & {} &
using Breit-Wigner forms \\ \hline $0.16$ & S.
Weinberg \cite{Weinberg} &
current algebra (non-linear $\sigma$-model) \\
\hline $0.20$ & J. Gasser,
H. Leutwyler \cite{Gasser} & one-loop
corrections, non-linear\\ {} & {}
& realization of chiral symmetry \\ \hline
$0.217$ & J. Bijnens et
al.\cite{Bijnens} & two-loop corrections, non-
linear\\ {} & {} &
realization of chiral symmetry  \\ \hline $0.26$
& M.K. Volkov
\cite{Volkov} & linear realization of chiral
symmetry \\ \hline $0.28$ &
A.N.Ivanov, N.Troitskaya \cite{Ivanov-Tr} & a
variant of chiral theory
with\\ {} & {} & linear realization of chiral
symmetry \\ \hline
\end{tabular} \label{tab:pipi.length} \end{table}

At first, let us remark about the result of Ref.\cite{Ishida} for the
$\pi\pi$ scattering length. We think that such a small value
(0.23 $m_{\pi^+}^{-1}$) has been obtained, because there has been used an
assumption about the negative $\pi\pi$ background in the phase shift.
Now, from Table~\ref{tab:pipi.length} we see that our results correspond to
the linear realization of chiral symmetry.

We have here presented model-independent results: the pole positions,
coupling constants and scattering lengths. Masses and widths of these states
that should be calculated from the obtained pole positions and coupling
constants are highly model-dependent. For instance, if we suppose that the
$f_0(665)$ is the $\sigma$-meson, then from the known relation~
$$g_{\sigma\pi\pi}=(m_\sigma^2-m_\pi^2)/\sqrt{2}f_{\pi^0}$$
(here $f_{\pi^0}$ is the constant of weak decay of the $\pi^0$:
$f_{\pi^0}=93.1$ MeV), we obtain ~$m_\sigma\approx 342$ MeV.

If we take the resonance part of amplitude in a non-relativistic form
(\cite{PDG-00}, p. 214)
$$T^{res}=\frac{\Gamma_{el}/2}{E_\sigma-E-i\Gamma_{tot}/2},$$
then we have $E_\sigma\approx 570\pm21$ MeV and $\Gamma_{tot}\approx
1400\pm30$ MeV; for the so-called relativistic form
$$T^{res}=\frac{\sqrt{s}\Gamma_{el}}{m_\sigma^2-s-i\sqrt{s}\Gamma_{tot}},$$
the following values are obtained: $m_\sigma\approx 850\pm20$ MeV and
$\Gamma\approx 1240\pm30$ MeV.

\section{Conclusions}

On the basis of a simultaneous description of the isoscalar $s$-wave channel
of the processes $\pi\pi\to\pi\pi,K\overline{K}$ with a parameterless
representation of the $\pi\pi$ background, a model-independent confirmation
of the $\sigma$-meson below 1 GeV is obtained. We emphasize that this is
a real evidence of this state, because we have not been enforced to construct
the $\pi\pi$ background.

A parameterless description of the $\pi\pi$ background is given only by
allowance for the left-hand branch-point in the proper uniformizing variable.
This seems to be related to the fact that the exchanges by nearest $\sigma$-
and $\rho$-mesons in the crossing channels contribute to the
$\pi\pi$-scattering amplitude with opposite signs (due to gauge invariance)
and compensate each other.

Note also that a light $\sigma$-meson is needed, for example, for an
explanation of $K\to\pi\pi$ transitions using the Dyson -- Schwinger model
\cite{M.Ivan} and for an explanation of the experimental value of the
pion-nucleon $\Sigma$-term ($\Sigma_{\pi N}\sim 40-70$ MeV) in a linear
$\sigma$-model based on the $U(3)\times U(3)$ quark effective Lagrangian
\cite{Nagy-Rus-Volk}.

Since all the fitted parameters in describing the $\pi\pi$ scattering are
only the positions of poles corresponding to resonances, we conclude that our
model-independent approach is a valuable tool for studying the realization
schemes of chiral symmetry. The existence of the low-lying state $f_0(665)$
with the properties of the $\sigma$-meson and the obtained
$\pi\pi$-scattering length ($a_0^0(\pi\pi)\approx 0.27 [m_{\pi^+}^{-1}]$)
suggest the linear realization of chiral symmetry.

The analysis of the used experimental data gives the evidence that the
${f_0}(980)$ and especially ${f_0}(1370)$ resonance (if exists --
variants 2 and 4), have the dominant $s{\bar s}$ component. Note that a
minimum scenario of the simultaneous description of processes
$\pi\pi\to\pi\pi,K\overline{K}$ goes without the ${f_0}(1370)$ resonance.
The best total $\chi^2/\mbox{N}_{DF}$ for both the analyzed processes is
obtained with the set of states: $f_0(665)$, $f_0(980)$, $f_0(1500)$ and
$f_0(1710)$.
The $K\overline{K}$ scattering length is very sensitive to whether the
$f_0(1370)$ state exists or not.

The $f_0(1500)$ has the approximately equal coupling constants with the
$\pi\pi$ and $K\overline{K}$ systems, which apparently could point up to its
dominant flavour-singlet ({\it e.g.}, glueball) component \cite{Amsler-Close}.

The $f_0(1710)$ is represented by the cluster corresponding to the state
with the dominant $s{\bar s}$ component. Although the lattice simulations
suggest that the lowest mass state of a pure glue would be the $0^{++}$ with
a mass of 1670$\pm$20 MeV \cite{lattice}, our result is in accord with
Refs.\cite{Volkov-Yudichev,Amsler-Close} where the $f_0(1500)$ has
been considered as a candidate for the scalar glueball. Note that QCD sum
rules \cite{Narison} and the K-matrix method \cite{Anisovich} showed both
the $f_0(1500)$ and $f_0(1710)$ are mixed states with large admixture of
the glueball component. Their conclusions about the glueball component
concord to our conclusion as to the $f_0(1500)$ but not $f_0(1710)$. It
seems that the complement of the combined consideration by the $\eta\eta$ and
$\eta\eta^{\prime}$ channels should not change substantially this situation
in view of the relative distance of the $\eta\eta$ threshold, and because the
glueball component is not coupled to the $\eta\eta^{\prime}$ system. Note
also that the conclusion of QCD sum rules \cite{Narison} about the existence
of light glueballs (below 1 GeV) contradicts the lattice calculations and
is not confirmed by our method.

We stress that our results are very decisive, because our approach is based
only on the first principles (analyticity-microcausality and unitarity),
immediately applied to the analysis of experimental data, and it is free from
dynamical assumptions, because a way of its realization is based on the
{\it mathematical} fact that a local behaviour of analytic functions
determined on the Riemann surface is governed by the nearest singularities
on all corresponding sheets. It is very important that we were able to
describe the considered coupled processes without diminishing the number of
fitted parameters by some dynamical assumptions.

We think that multichannel states are most adequately represented by
clusters, {\it i.e.}, by the poles on all the corresponding sheets.
Pole clusters give a main effect of resonances, and on the uniformization
plane they are their good representation.
The pole positions are rather stable characteristics for various models,
whereas masses and widths are very model-dependent for wide resonances.
Earlier one noted that the wide resonance parameters are largely controlled
by the nonresonant background (see, {\it e.g.} \cite{Achasov-Shest}). In part
this problem is removed by the parameterless and natural description of the
$\pi\pi$ background; there remains only a considerable dependence of resonance
masses and widths on the used model.
Therefore, for those states it is of a little sense to publish masses and
widths. It seems to be more right to publish the pole positions on all
corresponding sheets. To specify a pole cluster, we propose to use its
centre on the complex-energy plane (the real part of this centre).
We have made this for the $\sigma$-meson (variant 1) owing to its large
definition in the Particle Data Group tables.


\end{document}